\documentclass{PoS}
\usepackage{graphicx,epsfig}
\title{Exploring Dirac neutralinos and EW adjoint scalars of N=1/N=2 hybrid SUSY at colliders\thanks{Supported in parts by the Polish MNiSW Grants N N202 230337 and N N202 103838.}}

\ShortTitle{Exploring Dirac neutralinos and EW adjoint scalars at colliders}

\author{\speaker{Jan Kalinowski}\\
        Faculty of Physics, University of Warsaw, Hoza 69, 00681 Warsaw, Poland\\
        E-mail: \email{Jan.Kalinowski@fuw.edu.pl}}

\abstract{Properties of Dirac neutralinos and the corresponding EW scalar gauge bosons, as predicted by the N=1/N=2 hybrid supersymmetric model,  and prospects for their discovery at colliders are discussed.}

\FullConference{35th International Conference of High Energy Physics - ICHEP2010,\\
		July 22-28, 2010\\
		Paris France}

\begin{document}

\noindent {\bf 1. Introduction.}\\ 
In  the simplest N=1 supersymmetric extension of the Standard Model each particle is accompanied by a sparticle that differs in 
spin by half a unit. Thus the gaugino partners $\tilde{G}^\alpha_K$ of the gauge bosons $G^\mu_K$ are self-conjugate Majorana fields (with $K=C,I,Y$ for SU(3), SU(2) and U(1)). In non-minimal models additional gauginos $\tilde{G}'^\alpha_K$  together with the original N=1 gauginos $\tilde{G}^\mu_K$ can be combined to Dirac fields. For example, in ~\cite{us1,us2,us3}
the usual super-multiplets $\hat{\cal{G}}_K=\{\tilde{G}^\alpha_K,G_K^\mu\}$
are supplemented  by additional N=1 chiral super-multiplets
$\hat{\Sigma}_K=\{\sigma_K,\tilde{G}'^\alpha_K\}$
of extra gauginos $\tilde{G}'^\alpha_{K}$ and scalar partners
$\sigma_K$ in the adjoint representation of the corresponding gauge group 
to form a vector hyper-multiplet of N=2 SUSY~\cite{secondSusy}. To keep the theory chiral, masses of new matter multiplets are assumed heavy implying  an hybrid N=1/N=2 structure of the model. 

The transition from Majorana to Dirac fields renders the theory
[partially] $R$-symmetric 
with important consequences:  the baryon and lepton number violating operators are naturally suppressed and SUSY flavor-changing and CP-violating contributions 
are reduced significantly, widening the parameter space
for supersymmetric theories \cite{Rflav}.  The Dirac nature of gauginos alters significantly the collider phenomenology and has implications for the relic density of the Universe~\cite{dm}. Last, not least, since adjoint scalars $\sigma_K$ are $R$-parity even, they can be produced not only pairwise with large cross section
            but also singly in particle collisions making searches at colliders very interesting.  \\

\noindent {\bf 2. Dirac gauginos and adjoint scalars}\\
Apart from kinetic $\int d^4\theta\, \hat{\Sigma}^\dagger\, {\rm exp}[\hat G]\, \hat \Sigma$,  new terms in  the hybrid model appear:   the Majorana mass $M'$ for the new gauge superfields   
$\int d^2\theta M' \hat{\Sigma}\hat{\Sigma}$ as well as  the  Dirac mass $M^D$ coupling the original and new gauge superfields  $\int d^2\theta M^D\theta^\alpha \rm{tr} \hat{G}_{\alpha}\hat{\Sigma}$ (with gauge group indices understood), and 
new trilinear terms $W'_H=\int d\theta^2 \hat{H}_u\cdot(\lambda_ I \hat{\Sigma}_I   +\lambda_Y\hat{\Sigma}_Y)\hat{H}_d$ that couple new electroweak chiral to Higgs superfields.\footnote{  $M'$ and $M^D$ can be generated by contact terms, $\int d^4\theta \, {X^\dagger}\hat{\Sigma}\hat{\Sigma}$ and $\int d^2\theta \, {W'^\alpha}  W_{\alpha} \hat{\Sigma} $, when hidden sector spurion fields get vacuum expectation values, $\langle X\rangle =F\theta^2$ and $\langle W'^\alpha\rangle =D'\theta^\alpha$, respectively. }
Soft N=1  breaking terms complete the definition of the hybrid model.

New fields imply new couplings.  New gauginos $\tilde{G}'$  and sigma fields $\sigma$ couple minimally to the gauge fields, 
as required by the gauge symmetry. Also gauge symmetry dictates tree level couplings of the sigma fields to the gaugino  and sfermion-fermion fields. Fermion and  sfermion fields interact only with the standard gaugino
since only their mirror partners (assumed to be heavy) couple to $\tilde{G}'$. The tri-linear terms $W'_H$ in the superpotential imply Yukawa-type interactions of the EW sigma fields with higgsinos, while 
the Dirac gaugino mass generates, via the $D$-term, new tri- and quadri-linear  scalar couplings between $\sigma$ field(s) and sfermion pair 
$$
{\cal L}_{\sigma(\sigma)\tilde{f}\tilde{f}} = \left[ - \sqrt{2} g M^D (\sigma^a +\sigma^{a*})
            +i g^2 f^{abc} \sigma^{b*} \sigma^c \right]
     (\tilde{f}^\dagger_L T^a \tilde{f}_L -\tilde{f}^\dagger_R T^a \tilde{f}_R)$$
where $M^D$, $g$, $T$ and $f$ are the Dirac masses, couplings, generators and structure functions of the corresponding gauge group [note the minus sign in front of the quartic ${\tilde{f}}_R$ term, which 
corrects Eq.(2.16) in Ref.~\cite{us3}]. 
Since sfermions couple directly to gauge bosons and fermions, the above coupling, via sfermion/gaugino loops,  generates effective  couplings of $\sigma$ fields to gauge bosons and fermions. 
Note however that  $L$- and $R$-sfermion contributions
to the effective coupling come with opposite signs.  For colored $\sigma_C$ they cancel each other for mass-degenerate squarks.  In the EW sector only $L$ sfermions couple to $\sigma_I$, while for $\sigma_Y$ only partial cancellation occurs due to different hypercharges of $L$ and $R$ sfermions. The fermion-antifermion-sigma  coupling is also 
suppressed by the fermion mass, as evident from general chirality rules.

\begin{figure}[t]
\includegraphics[width=4.5cm,height=4.3cm]{neutmass.eps}\hskip 5mm
\includegraphics[width=4.5cm,height=4.3cm]{scalars.eps}\hskip 5mm
\includegraphics[width=4.5cm,height=4.3cm]{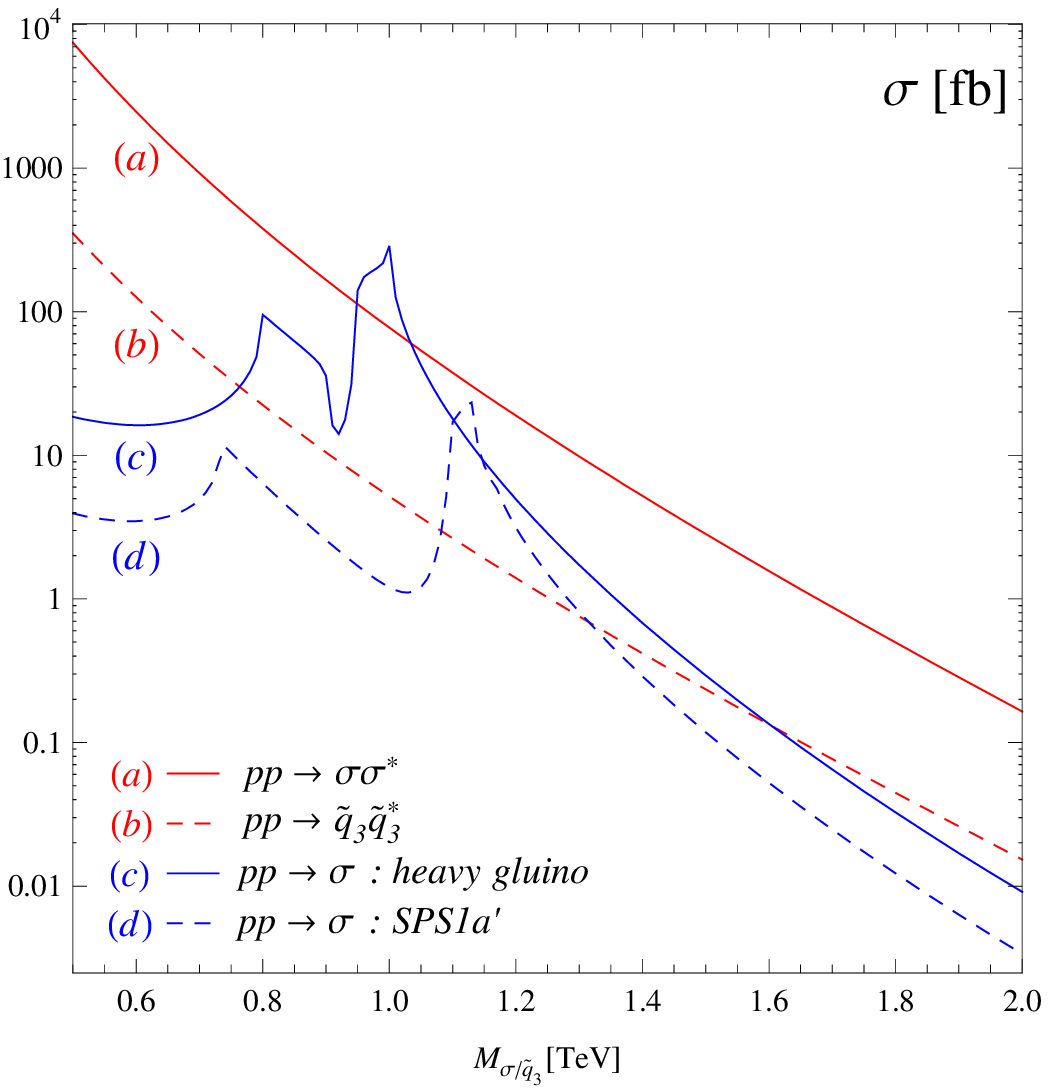}
\caption{Left: {\it Evolution of the neutralino masses as a function of the 
             parameter $y$ from the MSSM  to the Dirac case.} Middle: {\it  Neutral scalar masses as a function of the
           hypercharge soft scalar mass $m_Y$. } Right: { \it Cross sections for $\sigma_C$-pair (a), $\tilde{q}$-pair (b)  
         and single  $\sigma_C$ (c,d) production  
         at the LHC.  } }
\label{fig:one}
\end{figure}

In the $\tilde{G},\tilde{G}'$ basis the 2$\times$2 gaugino mass matrix has on the diagonal the soft SUSY breaking Majorana mass $M$ for the usual gaugino  and the supersymmetric Majorana $M'$ for the new one, and  on the off-diagonal  the Dirac mass  $M^D$. 
Its diagonalization  gives rise to two Majorana mass eigenstates, $\tilde{G}_1$ and $\tilde{G}_2$.   
There are two limiting cases of interest that can smoothly be connected by an auxiliary parameter $y\in [-1,0]$ setting $M'=y\, M^D/(1+y)$, $M=-y\, M^D$. Keeping $M^D=m_{\tilde{G}_1}$ fixed,  the standard  Majorana gaugino  is reached for $y=-1$ (with the other infinitely heavy), while for $y=0$ two degenerate Majorana states are combined to one Dirac 
${\tilde{G}}_D = \tilde{G} \oplus {\tilde{G}}'$.

In the electroweak sector this simple picture complicates once the gauge symmetry is broken. EW gauginos get coupled to higgsinos extending the chargino mass matrix to the 3$\times$3 and the neutralino to 6$\times$6. Nevertheless, for the  couplings $\lambda_I$ and $\lambda_Y$ as dictated by N=2 SUSY, in the limit $y=0$ three Dirac neutralinos are obtained as illustrated in Fig. \ref{fig:one}(left).  

The scalar electroweak sector, apart from the usual Higgses, includes also the iso-triplet and hypercharge fields. Expanding the fields around $vev$'s and eliminating the Goldstone modes, and diagonalising the mass matrices one ends up with 3 parity-odd neutral $a_i$, 4 parity-even neutral $s_i$ and 3 charged scalar $s_i^\pm$ mass-eigenstates.  The small value of the iso-triplet $vev$, $v_I\leq 3$ GeV, to comply with constraints on the $\rho$ parameter, can be met assuming the soft mass of $\sigma_I$ of order TeV. On the other hand, the hypercharge $v_Y$ and the $\sigma_Y$ soft mass can be arbitrary. Fig.\ref{fig:one}(middle) illustrates the masses of the four parity-even eigenstates as a function of soft mass parameter $m_Y$. \\

\noindent{\bf 3. Collider signatures} \\
The phenomenology of the hybrid model is very rich. Here we show only a few examples; more can be found in~\cite{us1,us2,us3}.

Since sgluons $\sigma_C$, scalar partners of gluons,  have a large color charge they might be more copiously produced than squarks at the LHC. In addition they can be produced singly as an $s$-channel resonance in $gg$ fusion via the loop induced coupling. Fig.\ref{fig:one}(right) shows the expected cross section at the LHC in specific scenarios considered in ref.\cite{us1}. Such processes would be really spectacular given the fact that the $\sigma$ decays  give rise to signatures with multi-jets and large $E_{miss}$ that should easily be detectable. 

\begin{figure}[t]
\includegraphics[width=4.5cm,height=4.3cm]{nn_prod_hybrid.eps}\hskip 5mm
\includegraphics[width=4.8cm,height=4.3cm]{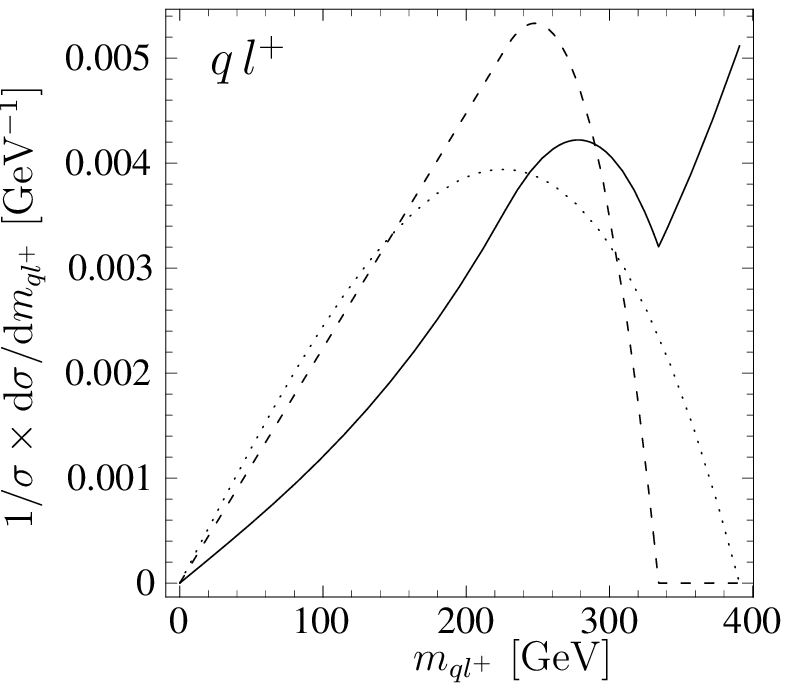}\hskip 5mm
\includegraphics[width=4.2cm,height=4.3cm]{gg2sa.eps}
\caption{Left: {\it  Cross sections for pair
         production of wino-like neutralinos near threshold in the MSSM and
         the Dirac theory.} Middle: {\it $ql^+$ invariant mass distributions for squark decay chains involving
             Majorana or Dirac neutralinos: in the MSSM $\tilde{q}$ and $\tilde{q}^*$ 
             decay chains lead to identical distributions (solid),
             in contrast to the Dirac case (dashes and dots, respectively).}
 Right: {\it Cross sections
             $\sigma_0(\gamma\gamma\to\phi)$ for $\phi=s_{Y,I}$ and
             $a_{Y,I}$.  }   }
\label{fig:two}
\end{figure}

Lepton colliders are ideal laboratories to study the properties of EW gauginos. The phenomenology of Dirac gauginos is characteristically different from Majorana. 
First, Dirac pair production is not forward-backward symmetric. Second, near threshold the diagonal neutralino pair is excited in $P$-waves for Majorana while in $S$-waves for Dirac, see Fig.\ref{fig:two}(left). The enhanced production for Dirac neutralinos may extend the search reach at colliders. By the same token,  the neutralino  annihilation cross section in the MSSM shows a $P$-wave, while in the Dirac case only an $S$-wave suppression into a fermion and anti-fermion pair and non-vanishing even for massless fermions.  
Thus, if the lightest neutralino is  the LSP and stable, its nature has important consequences for cold dark matter phenomenology~\cite{dm}. 
At hadron colliders the nature of the neutralino can be tested in the celebrated cascade squark decay. The decay chains differ:  $\tilde{q}_L \to q \, \tilde{\chi}^0_2
    \to q \, l^\pm \, \tilde{l}^\mp_R \to q \, l^\pm \, l^\mp \, \tilde{\chi}^0_1$ for MSSM, 
 $ \tilde{q}_L \to q \, \tilde{\chi}^{c0}_{D2}
    \to q \, l^+ \, \tilde{l}^-_R
    \to q \, l^+ \, l^- \, \tilde{\chi}^0_1$ for Dirac, which will leave  a
characteristic imprint on the angular distributions of visible decay jets and
leptons, as seen in  Fig.~\ref{fig:two}(middle) for $SPS1a'$-like scenario.
 
As a final example we consider the production processes of EW adjoint scalars.  Since $Q$, $I_3$ and $Y$ of  neutral scalars vanish, they cannot be pair-produced at $e^+e^-$. However, the charged pairs  $\sigma^+_{I}\sigma^-_{I}$ can be excited through $\gamma,Z$ exchanges with cross sections exceeding the charged Higgs states due to their larger EW charge. 
Resonant $s$-channel formation in fermion-antifermion annihilation is strongly suppressed since the tree level couplings are suppressed by the fermion mass. However, they can be excited in $\gamma\gamma$. Adopting the SPS1a' scenario, the reduced $\sigma_0(\gamma\gamma\to \phi)$ cross section (the luminosity factor removed) can be of order 1 fb, see Fig.2(right).\\

\noindent{\bf Summary} \\
In extended SUSY models, the gauginos can be of Majorana or Dirac type. Their nature can be tested at the LHC and linear colliders in a variety of production and decay channels. In addition, the accompanying adjoint scalars make the collider phenomenology very exciting.
\\

\noindent{\bf Acknowledgments} ~ It is my pleasure to thank S.Y. Choi, D. Choudhury, M. Drees, A. Freitas, J.M. Kim, E. Popenda and Peter Zerwas for a fruitful collaboration.

\end{document}